\documentclass{article}
\usepackage[T1]{fontenc}
\usepackage{newunicodechar}
\newunicodechar{–}{\textendash}
\PassOptionsToPackage{table}{xcolor}
\usepackage{iclr2027_conference,times}

\usepackage[hidelinks]{hyperref}
\usepackage{url}
\usepackage{graphicx}
\usepackage{xspace}
\usepackage{booktabs}
\usepackage{multirow}
\usepackage{capt-of}
\usepackage{wrapfig}
\usepackage{subcaption}
\usepackage{placeins}
\usepackage{float}
\usepackage[bottom]{footmisc}

\newcommand{\approach}{SWE-at-Scale\xspace}
\xspaceaddexceptions{+}

\title{\approach: Scalable Issue Synthesis from Historical Bug Fixes}

\author{Silin Chen\textsuperscript{1}, Zihan Zhang\textsuperscript{1},
Xiaodong Gu\textsuperscript{1}\thanks{Corresponding author.},
Jiale Huang\textsuperscript{2}, Yuling Shi\textsuperscript{1}\\
\bfseries Han Li\textsuperscript{1}, Zhao Shan\textsuperscript{3},
Chengcheng Wan\textsuperscript{4,6}, Zhensu Sun\textsuperscript{5},
Haibing Guan\textsuperscript{1}\\[0.5em]
\normalfont\small
\textsuperscript{1}Shanghai Jiao Tong University \quad
\textsuperscript{2}North China Electric Power University \quad
\textsuperscript{3}Tsinghua University\\
\normalfont\small\textsuperscript{4}East China Normal University \quad
\textsuperscript{5}Singapore Management University \quad
\textsuperscript{6}Shanghai Innovation Institute}

\iclrfinalcopy
\begin{document}

\setcounter{footnote}{1}
\maketitle
\fancyhead[L]{Preprint}

\begin{abstract}
    As coding agents become increasingly capable, robust evaluation requires a broader and more diverse supply of verifiable software engineering (SWE) tasks. However, real-world tasks are constrained by the natural frequency of issue reporting and the substantial cost of manual curation. Furthermore, existing synthetic approaches rely primarily on the pretrained knowledge of large models coupled with heuristic prompting, often failing to capture the authentic, complex mechanics of software failures. To exponentially scale SWE task production, we introduce \approach, a framework that automatically synthesizes realistic issues by reverse-engineering real-world software fixes. Our key insight is that historical fixes encapsulate deep mechanistic knowledge of bug formation: inverting a fix transforms repair supervision into bug-induction supervision grounded in actual development scenarios. \approach transfers these learned bug patterns to novel code contexts, repeatedly sampling diverse bug typologies to construct executable SWE tasks. Extensive experiments show that \approach generates more diverse and realistic SWE tasks than current synthesis baselines while posing greater challenges to state-of-the-art coding agents. Notably, \approach can generate distinct bug types from a single code context, with bug diversity scaling robustly even up to 200 generations per file. Overall, \approach turns finite real-world fixes into reusable supervision for scalable synthesis of realistic, diverse, and challenging SWE tasks.
\end{abstract}

\begin{figure}[!b]
    \centering
    \includegraphics[width=0.94\linewidth]{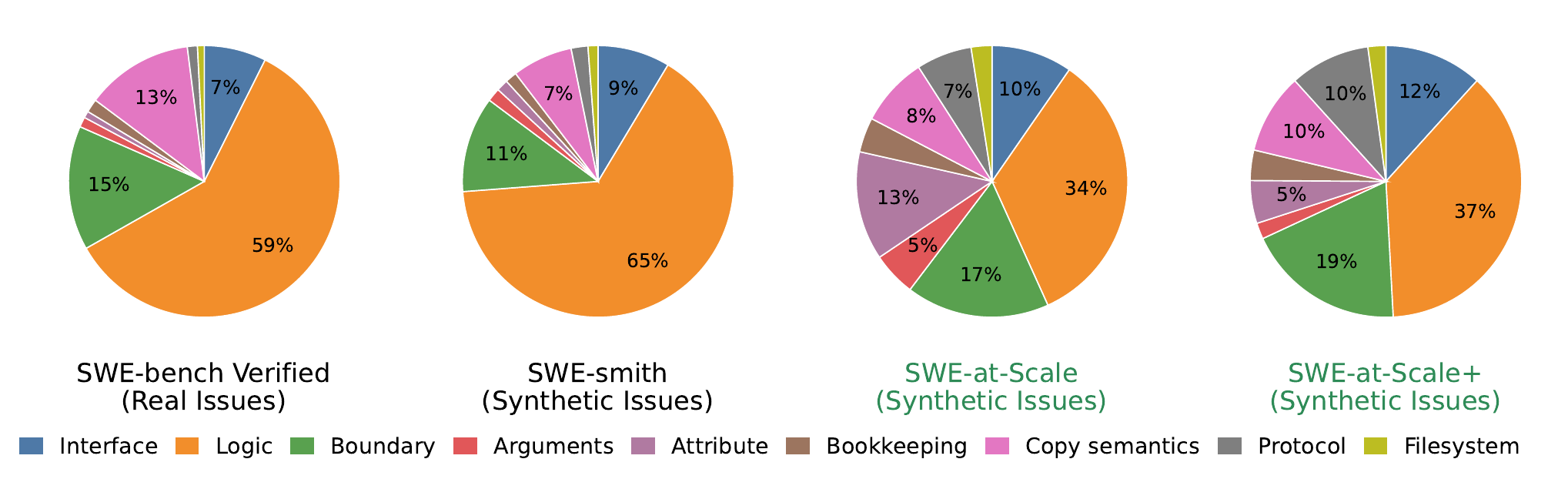}
    \caption{Bug type distributions for real-world SWE-bench Verified tasks, SWE-smith synthetic tasks, and \approach synthetic SWE tasks.}
    \label{fig:bug-type-distribution}
\end{figure}

\section{Introduction}

Repository-level software engineering (SWE) tasks have become a central testbed for evaluating coding agents~\citep{chen2026sweexpexperiencedrivensoftwareissue,lin2024llmscontinuouslearnersimproving,ma2026llmagentscoderepositories,chen2026skillforgeselfdistillingagentsprojectspecific,chen2026repo0designdrivenzerotoallcode,lin2026knowfixqadrivenrepository, gu2025effectiveness, shi2026earlyevalcheaperagentevaluation,10.1109/ICSE55347.2025.00204}. A useful SWE task involves more than a code change~\citep{yang2025swesmith,sonwane2025bugpilotcomplexbuggeneration}: it should preserve a realistic repository context, introduce an observable behavioral bug, provide a meaningful issue description, and remain executable within the repository's
testing infrastructure.

Existing real-world SWE benchmarks~\citep{jimenez2024swebenchlanguagemodelsresolve} are inherently difficult to scale, as new tasks can only emerge at the pace of real software development~\citep{zhang2025swebenchgoeslive}.
Collecting these tasks is further constrained by repository availability and the cost of constructing executable environments~\citep{jain2025r2egymproceduralenvironmentshybrid}. Consequently, scaling SWE task data solely through naturally occurring issues requires continuously discovering and validating additional real-world development events.

Recent work has explored large language models (LLMs) as a means of scaling SWE task construction. SWE-smith~\citep{yang2025swesmith} leverages LLMs to inject bugs by prompting them to modify selected code entities, enabling the synthesis of large numbers of synthetic task instances. SWE-Synth~\citep{pham2025swesynthsynthesizingverifiablebugfix} employs LLM-based component re-implementation followed by test-based filtering, while BugPilot~\citep{sonwane2025bugpilotcomplexbuggeneration} explores feature addition and behavioral regression as a source of synthetic bugs. These approaches demonstrate that LLMs can generate bugs beyond simple hand-designed
mutation rules.
However, without dedicated training for bug injection, these approaches largely depend on the bug-generation knowledge already encoded in general-purpose LLMs and elicited through prompting, making repeated generation more likely to revisit familiar bug patterns rather than continually expanding the diversity of synthesized tasks. This motivates a different scaling strategy: rather than repeatedly prompting a general-purpose LLM, we reverse real-world bug fixes to construct correct-to-buggy supervision and train a reusable bug-injection model that can generate diverse bugs in new code contexts.

We introduce \approach, a framework that turns real software fixes into a scalable source of repository-level SWE tasks. Rather than treating each historical fix as a single, one-off task instance, \approach reverses bug-fixing transformations and uses them as supervision for learning how bugs can be induced. Aggregating such reversed transformations across diverse repositories allows the model to learn a reusable distribution of real-world bug patterns, instead of relying solely on the bug-generation knowledge already encoded in general LLMs. Through this reverse-learning formulation, \approach can repeatedly sample these real-world bug patterns and instantiate them in new code contexts, allowing a finite collection of real fixes to support the synthesis of a much larger and more diverse set of executable SWE tasks.

\begin{figure*}[t]
    \centering
    \includegraphics[width=\textwidth]{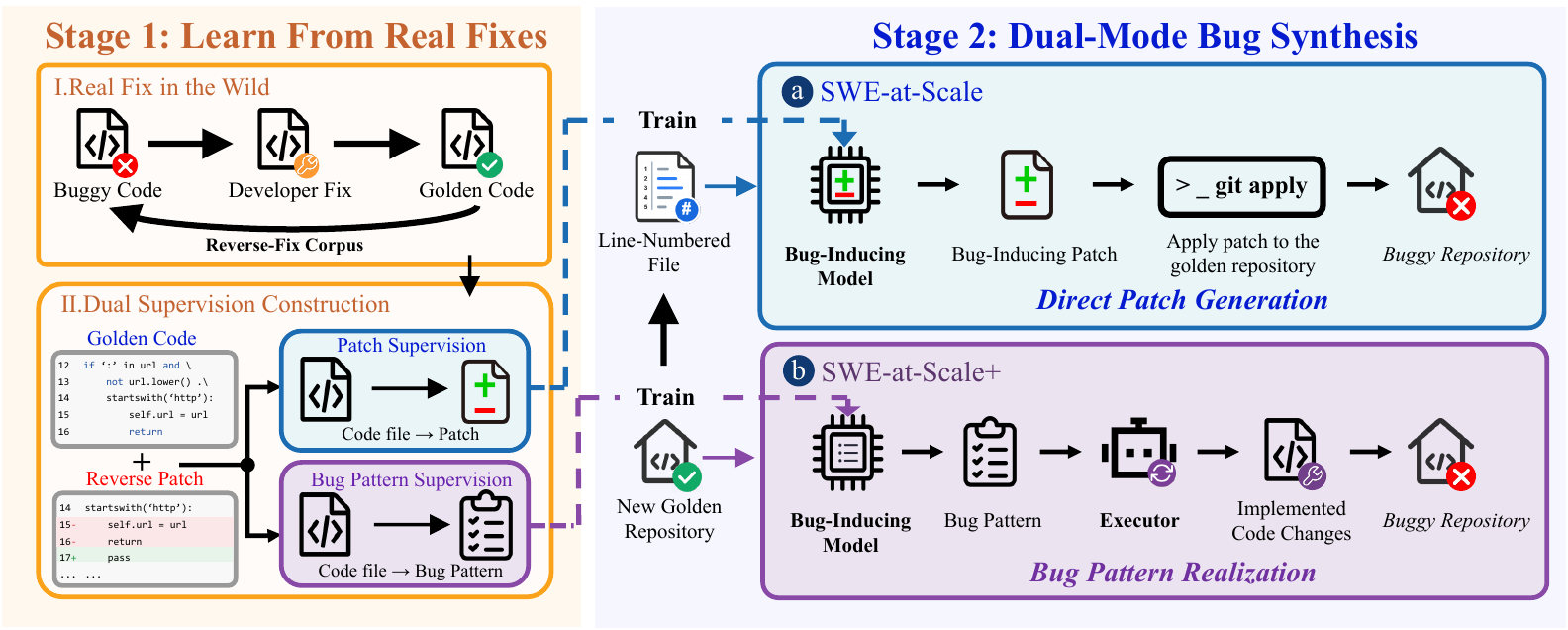}
    \caption{Overview of \approach. Stage 1 reverses real software fixes to construct bug-inducing patch and bug-pattern supervision, which is used to train the synthesis models. Stage 2 applies the learned transformations to new golden repositories through two complementary synthesis models.}
    \label{fig:framework}
    \vspace{-8pt}
 \end{figure*}

We train \approach and \approach+ on 13,407 real-fix transformations collected from 688 repositories, using Qwen3.5-9B as the backbone model. \approach directly learns to generate bug-inducing patches, while \approach+ explicitly learns natural-language bug patterns that describe how the target code should be modified to induce a bug.
Experiments show that \approach synthesizes SWE tasks that are realistic, challenging, diverse, and scalable under repeated generation. In pairwise comparisons with real-world SWE-bench Verified issues, human evaluators select \approach- and \approach+-generated bugs as more likely to occur in real software development in 55.9\% and 61.2\% of cases, respectively. The synthesized tasks also remain challenging for current coding agents: across four resolution models, \approach achieves Pass@1 between 31.68\% and 48.48\%, while \approach+ achieves between 16.81\% and 37.02\%, with \approach+ approaching the difficulty of human-expert-designed tasks such as DeepSWE~\citep{huang2026deepswemeasuringfrontiercoding}. As shown in Figure~\ref{fig:bug-type-distribution}, \approach covers a more balanced range of bug types than existing synthesis methods, suggesting that learning from real fixes captures a richer bug distribution than relying on general LLMs' knowledge alone. This diversity also persists under repeated sampling: even after 200 generations per file, 88.27\% of usable patches remain unique, indicating that the generated bugs do not quickly collapse to repeated outcomes.

Our contributions are threefold:
1)~We introduce \approach, a scalable paradigm for SWE task synthesis. \approach reverses real bug fixes as supervision for learning how bugs are introduced, allowing bug patterns observed in real development to be instantiated in new code contexts;
2)~We construct a large-scale training corpus of 13,407 real-fix transformations from 688 repositories and develop two synthesis variants: \approach directly learns bug-inducing patches, while \approach+ externalizes bug-inducing logic as natural-language patterns for more explicit realization;
and
3)~Through controlled comparisons with real-world and synthetic SWE tasks, we show that \approach produces realistic, diverse, and challenging executable tasks, and that its bug diversity continues to increase as the inference-time sampling budget grows.

\section{\approach: Software Task Synthesis from Real Fixes}
The core principle of \approach is to learn how bugs are introduced by reversing real software fixes. The learned bug patterns can be adapted to new code context, enabling the synthesis of large-scale SWE tasks. The pipeline involves two stages, as illustrated in Figure~\ref{fig:framework}. Given a corpus of real bug-fix commits, \approach constructs reversed (code, buggy code) pairs (Stage I). A bug inducing model is trained using the collected pairs (Stage II). The trained model automatically generates buggy code given arbitrary source code. \approach learns these transformations and uses them to synthesize new SWE task instances.

\subsection{Training Data Collection}
\begin{figure*}[t]
    \centering
    \begin{minipage}[t]{0.48\textwidth}
        \centering
        \includegraphics[width=\linewidth]{
            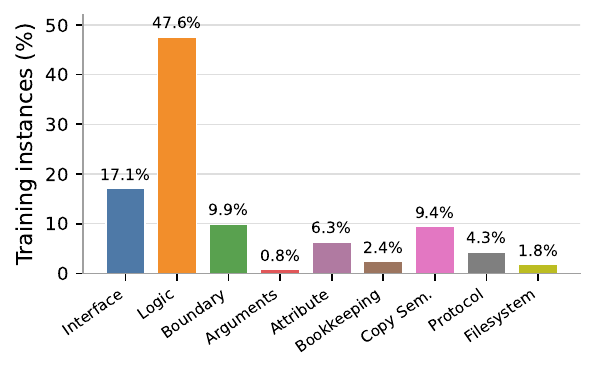
        }
        \captionof{figure}{Bug type distribution in the training dataset.}
        \label{fig:training-bug-types}
    \end{minipage}
    \hfill
    \begin{minipage}[t]{0.48\textwidth}
        \centering
        \includegraphics[width=\linewidth]{
            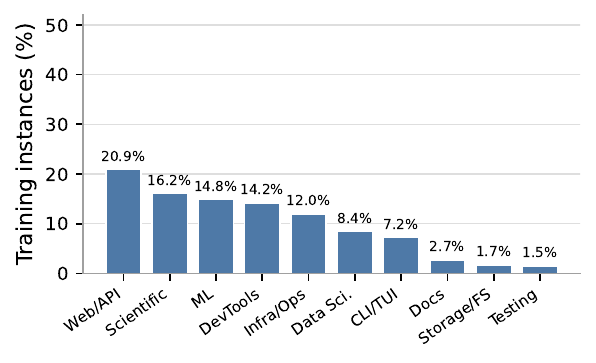
        }
        \captionof{figure}{Distribution of training instances across repository
        domain categories.}
        \label{fig:repository-domains}
    \end{minipage}
    \vspace{-16pt}
\end{figure*}

To train the bug-inducing model, we construct a corpus of historical bug fixes and reverse them into (golden code, buggy code) pairs. We collect bug-fix instances from SWE-Fixer~\citep{xie-etal-2025-swe}, SWE-rebench~\citep{badertdinov2025swerebenchautomatedpipelinetask},
SWE-bench-Live~\citep{zhang2025swebenchgoeslive}, and
SWE-Gym~\citep{pan2025trainingsoftwareengineeringagents}. The four sources contain more than 134K issue--fix instances in total. For each instance, we identify the source files modified by the real fix and recover their versions before and after the fix. For each real-fix instance, we first apply the original golden patch to the buggy repository to reconstruct the corresponding fixed source files. We then compute a new unified diff from each fixed file back to its original buggy version. This reverse-direction diff serves as the bug-inducing patch used for training \approach. To avoid repository-level overlap with our evaluation data, we exclude all instances from repositories represented in SWE-bench Verified.

We apply a series of quality filters to the collected records. We remove examples with missing or unparseable patches, non-code files, or changes dominated by imports, comments, whitespace, or formatting. We further discard structurally incomplete patches that do not represent a complete and internally consistent code transformation. For each retained record, we reconstruct the buggy and fixed file states, generate unified diffs in both directions, and verify that the forward and reverse patches apply successfully and reproduce the corresponding original states through round-trip application. Following BugPilot~\citep{sonwane2025bugpilotcomplexbuggeneration}, we annotate the remaining examples with bug types and remove examples whose serialized input--output sequences exceed the 64K-token limit.

From the remaining candidates, we construct the final training dataset through distribution-aware sampling, aiming to balance bug types, repositories, and context lengths as much as permitted by the available real-fix data. We further manually inspect the selected patches to ensure their quality. After filtering, reconstruction, distribution-aware sampling, and manual inspection, the final training dataset contains 13,407 real-fix transformations from 688 repositories, which provide the training supervision for both \approach synthesis models.

Figures~\ref{fig:training-bug-types} and~\ref{fig:repository-domains} summarize the resulting data distribution. Our balancing strategy substantially broadens coverage across bug types, repositories, and software domains while preserving the naturally occurring characteristics of real software fixes. Logic and Interface transformations remain the most frequent categories, reflecting their prevalence in the real-fix data rather than an intentional sampling bias.

\subsection{Training Bug-Inducing Model}

\approach uses reversed real-fix transformations as supervision for training the bug inducing model. For each real-fix transformation, the fixed file serves as the golden input, while the reversed fix serves as the bug-inducing target. We use Qwen3.5-9B~\citep{qwen3.5} as the backbone model and fine-tune it to generate a unified diff from the complete golden source file annotated with line numbers. The generated diff transforms the correct code into its corresponding buggy state. Providing the complete file allows the model to infer the relevant edit locations from the surrounding code context rather than receiving predefined locations, thereby learning jointly where to modify the code and how to realize the corresponding bug-inducing changes. Unlike synthesis with general LLMs without dedicated bug-inducing training, this formulation explicitly trains the model on correct-to-buggy transformations observed in real software development.

At synthesis time, \approach can operate on any source code file in a correct repository state and generate a candidate bug-inducing patch, which is applied to the corresponding repository using \texttt{git apply}. Successfully applied patches produce candidate buggy repositories that are subsequently passed to execution-based validation. Sampling temperature controls how broadly \approach explores the learned bug-inducing distribution, enabling repeated sampling from the same code context for scalable task synthesis.

Direct patch generation provides a simple and efficient way to learn bug-inducing patches from reversed real-fix transformations, but it requires the model to jointly determine what bug to introduce and how to realize it directly in code. The underlying bug pattern therefore remains implicit in the generated patch. To study whether making this pattern explicit can provide stronger guidance for bug synthesis, we further introduce an enhanced variant, \approach+, which represents the intended bug transformation in natural language before realizing it in code.
To construct supervision for \approach+, we use DeepSeek-V4-Flash-0731~\citep{deepseekai2026deepseekv4} as an expert teacher. For each verified inverse transformation, the teacher is prompted to infer an executor-ready natural-language bug pattern that specifies the task category, target location, current behavior, specific code modification, implementation constraints, and expected observable effect. Using the same golden file as input, \approach+ is trained to predict this pattern instead of the unified-diff target used by \approach. At synthesis time, the generated bug pattern and golden code are provided to an executor that realizes the specified modifications. To minimize the influence of the executor's prior knowledge of bug generation, we present the pattern as ordinary code-modification instructions without revealing that the requested changes are intended to introduce a bug. The resulting code forms the candidate buggy repository for subsequent execution-based validation.

\subsection{Task Assembly and Validation}
Finally, \approach assembles SWE tasks using the generated code changes. Following SWE-smith~\citep{yang2025swesmith} and BugPilot~\citep{sonwane2025bugpilotcomplexbuggeneration}, we construct the final task through execution-based validation and issue-description generation. We first apply the generated code changes to the corresponding golden repository and execute its test suite. We retain only candidates that cause at least one previously passing test to fail, referred to as a Pass-to-Fail (P2F) test.
For each validated buggy repository, we generate a user-facing issue description without exposing the synthesis process. We provide an LLM with the generated code diff, the source code of a P2F test, its execution output and two demonstrations~\citep{yang2025swesmith}, and prompt it to produce a GitHub-style issue describing the observed and expected behavior. The resulting issue description is paired with the validated buggy repository to form the final SWE task instance.

\section{Experiments}
\subsection{Setup}

\paragraph{Models.}
For baseline task synthesis methods, we use DeepSeek-V4-Flash-0731~\citep{deepseekai2026deepseekv4}, GPT-5.4-mini, and Qwen3.5-9B~\citep{qwen3.5} as the underlying LLMs. For \approach+, we use DeepSeek-V4-Flash-0731 as the expert model for generating natural-language bug patterns, and Qwen3.5-9B as the base model for training both \approach and \approach+. The prompts are provided in our anonymous repository\footnote{\url{https://anonymous.4open.science/r/SWE-at-scale-FA0A}}. For downstream issue resolution, we use DeepSeek-V4-Flash-0731 (DS-V4-Flash), GPT-5.4-mini, GLM-5.3-Flash~\citep{glm5team2026glm5vibecodingagentic}, and Qwen3.8-Flash-Next (Qwen3.8-Flash)~\citep{qwen3.8flashnext} as the coding agents to evaluate the difficulty of the synthesized task instances. 

\paragraph{Baselines.}
We compare \approach against three representative synthesis methods: SWE-smith~\citep{yang2025swesmith} with its LLM-Modify and LLM-Rewrite strategies, and BugPilot~\citep{sonwane2025bugpilotcomplexbuggeneration} with its FeatAdd strategy. We also include SWE-bench Verified~\citep{jimenez2024swebenchlanguagemodelsresolve} and DeepSWE~\citep{huang2026deepswemeasuringfrontiercoding} as a reference set of real-world SWE issues.

\paragraph{Implementation Details.}
We fine-tune \approach and \approach+ on the 13,407 real-fix transformations
for one epoch using supervised fine-tuning (SFT) with LoRA~\citep{hu2021loralowrankadaptationlarge} and a 64K-token context window.
All training is performed on a single NVIDIA RTX PRO 6000 GPU; the complete LoRA and optimization configuration is provided in
Appendix Table~\ref{tab:lora-config}.

For task synthesis, each method receives up to five attempts per target file:
the first uses temperature 0 and the remaining four use temperature 1.0.
Generation terminates once a candidate causes at least one previously passing regression test to fail; otherwise, the instance is discarded after five attempts. For controlled comparison, all methods synthesize bugs from the same target file and golden repository state associated with each SWE-bench Verified instance. We use the default FeatAdd configuration from BugPilot. For SWE-smith, both LLM-Modify and LLM-Rewrite receive the complete target file and are constrained to the smallest enclosing functions or classes containing the golden-patch hunks. LLM-Rewrite masks and reimplements these units, whereas LLM-Modify keeps them visible and may introduce any number of modifications within them.

We evaluate synthesized tasks using mini-SWE-agent~\citep{yang2024sweagent} with temperature 0 and a maximum of 250 actions. To avoid model-specific generation--resolution bias, we use model-disjoint evaluation: a model never resolves tasks whose problem statements were generated by itself. Each experimental configuration is run independently three times, and we report the mean across runs.

\begin{table*}[t]
\centering
\small
\setlength{\tabcolsep}{5pt}
\begin{tabular}{cc|cccc}
\toprule
\multirow{1}{*}{Method}
& \multirow{1}{*}{Developer}
& \multicolumn{1}{c}{DS-V4-Flash}
& \multicolumn{1}{c}{GPT-5.4-mini}
& \multicolumn{1}{c}{GLM-5.3-Flash}
& \multicolumn{1}{c}{Qwen3.8-Flash} \\
\midrule

\rowcolor{gray!15}
\multicolumn{6}{c}{\textit{Real-world Tasks}} \\

DeepSWE
& Human & 54.40 & 24.30 & 63.40 & 58.70 \\

SWE-bench Verified
& Human & 91.80 & 69.60 & 94.00 & 97.00 \\

\midrule

\rowcolor{gray!15}
\multicolumn{6}{c}{\textit{Synthetic Tasks}} \\

\multirow{3}{*}{LLM-Modify}
& DS-V4-Flash & 79.85 & 35.82 & 57.78 & 63.96 \\
& GPT-5.4-mini & 75.17 & 56.04 & 59.06 & 66.78 \\
& Qwen3.5-9B & 77.84 & 40.21 & 51.55 & 64.86 \\
\cmidrule(lr){1-6}

\multirow{3}{*}{LLM-Rewrite}
& DS-V4-Flash & 67.80 & 25.14 & 47.36 & 56.32 \\
& GPT-5.4-mini & 60.66 & 31.86 & 45.15 & 55.93 \\
& Qwen3.5-9B & 70.90 & 38.52 & 48.36 & 72.84 \\
\cmidrule(lr){1-6}

FeatAdd
& DS-V4-Flash & 58.62 & 37.93 & 62.07 & 62.07 \\

\midrule

\multicolumn{2}{c|}{\approach} 
& 48.48 & 31.68 & 42.15 & 44.44 \\

\multicolumn{2}{c|}{\approach+} 
& \textbf{33.19} & \textbf{16.81} & \textbf{35.32} & \textbf{37.02} \\

\bottomrule
\end{tabular}

\caption{Downstream issue resolution performance (\%) across real-world and synthesized SWE task instances. \textit{Developer} denotes the model or human source that constructed each task instance.}
\label{tab:task-difficulty}
\vspace{-16pt}
\end{table*}

\subsection{Evaluation Metrics}

We evaluate synthesized task instances in terms of realism, usability, and difficulty. We measure realism using the \textbf{Human Preference Rate (HPR)}, following the human evaluation protocol of CodeOCR~\citep{shi2026seeingcodingeffectivenessvision}. Three software engineers with at least three years of experience compare each synthesized issue--patch pair with a real-world counterpart within the same context and judge which one they would be more likely to encounter in real-world software development; the final judgment is determined by majority vote. We measure usability using the \textbf{Usability Rate}, defined as the proportion of target base commits for which at least one generated task instance passes our execution and quality filters across up to five synthesis attempts. An instance must cause at least one previously passing regression test to fail, while its corresponding code changes are also manually judged to be non-trivial rather than simple modifications such as replacing code with \texttt{pass}, \texttt{TODO}, or \texttt{return None}. Finally, we use \textbf{Pass@1} to measure task difficulty based on the success rate of downstream issue resolution.

\section{Results}

\subsection{Synthesis Quality and Task Difficulty}

\begin{wrapfigure}{r}{0.48\textwidth}
   \centering
   \vspace{-6pt}
   \includegraphics[width=\linewidth]{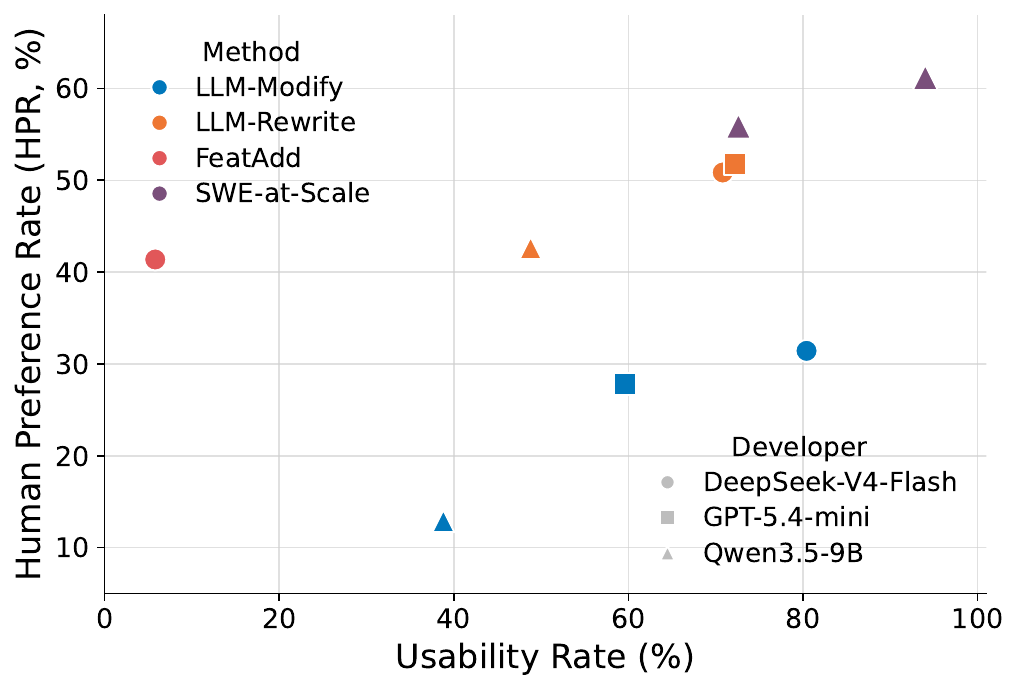}
   \caption{Usability Rate versus Human Preference Rate (HPR).}
   \label{fig:synthesis-quality-scatter}
   \vspace{-8pt}
\end{wrapfigure}

\textbf{\approach synthesizes more usable and realistic SWE tasks.}
Figure~\ref{fig:synthesis-quality-scatter} compares the Usability Rate and HPR of different synthesis methods. Both \approach variants occupy a more favorable region of the usability--realism space than existing synthetic baselines.
\approach+ achieves the strongest joint performance, with a 94.0\% Usability Rate and a 61.2\% HPR, while \approach achieves 72.6\% and 55.9\%, respectively. In comparison, the best LLM-Rewrite configuration reaches 72.2\% usability and 51.8\% HPR, while LLM-Modify exhibits substantially lower human preference across generation models. The gap between the two baselines is consistent with their construction procedures. LLM-Modify explicitly prompts an LLM to edit selected code locations to inject an issue, which can produce localized mutations with little connection to how defects arise during ordinary development. LLM-Rewrite instead asks the model to reimplement existing code entities, so an incorrect implementation can emerge as an unintended behavioral regression, a process that more closely resembles the accidental introduction of bugs by human engineers. FeatAdd similarly achieves high realism by deriving bugs from regressions that arise during feature development; however, the open-ended nature of feature implementation provides weak control over whether a usable regression will actually emerge, resulting in substantially lower bug-inducing usability.
We attribute the stronger performance of \approach+ to its explicit representation of bug logic in natural language. Rather than requiring the model to implicitly infer bug-inducing logic from code transformations, \approach+ externalizes this logic as a natural-language bug pattern, providing a more explicit learning signal that aligns well with the language-modeling capabilities of LLMs. These results suggest that learning bug-inducing transformations by reversing real-fix transformations captures bug patterns that better reflect real development practice than mechanisms that explicitly prompt an LLM to modify or rewrite code to naturally inject an issue.

\textbf{\approach produces synthetic bugs whose difficulty is competitive with human-expert-designed SWE tasks.}
As shown in Table~\ref{tab:task-difficulty}, both \approach variants consistently yield lower Pass@1 than SWE-bench Verified and existing synthetic baselines across the evaluated resolution models, with \approach+ achieving the lowest Pass@1 in all settings. This consistent trend indicates that the
observed difficulty generalizes across different resolution models rather than being specific to a particular solver. Notably, \approach-generated tasks also approach the difficulty of DeepSWE, whose tasks are designed by human experts.
This result provides further evidence that supervision derived from reversed real-fix transformations can teach models to construct challenging SWE tasks that approach the difficulty of expert-designed ones.

\subsection{Scaling Bug Diversity at Training and Inference Time}

We study three complementary factors that shape the diversity of
\approach-generated bugs: \textbf{repository coverage},
\textbf{sampling temperature}, and \textbf{sampling budget}. Repository coverage determines the diversity of bug-fixing patterns exposed to the model during training. At inference time, sampling temperature controls how broadly the model explores the learned bug-inducing distribution for a given code
context, while sampling budget determines how extensively that distribution is sampled. Together, these factors allow us to examine how bug diversity scales through both training-time data coverage and inference-time sampling.

\textbf{Mean Bug Type Entropy.} We measure how many different types of bugs \approach can generate for a single file. This differs from counting bug types over the whole dataset, where diversity can come simply from using many different files. To this end, we treat each target file as a fixed code context and quantify the diversity of its generated candidates using \textit{Mean Bug Type Entropy} (MBTE)~\citep{chen2026quantitytrajectorydiversityscaling,garg2026infosynthinformationguidedbenchmarksynthesis}. For target file $i$ under setting $s$, we compute its entropy as $H_{i,s}=-\sum_k p_{i,s}(k)\log_2 p_{i,s}(k)$, where $p_{i,s}(k)$ denotes the proportion of usable candidates assigned to bug type $k$. We then macro-average across the $N$ target files and report $\mathrm{MBTE}_s=\frac{1}{N}\sum_{i=1}^{N}H_{i,s}$. Higher MBTE therefore indicates that, on average, the model can generate a broader distribution of bug types from the same code context.

\begin{figure*}[t]
    \centering
    \begin{minipage}[t]{0.40\textwidth}
        \centering
        \includegraphics[height=2.0in]{
            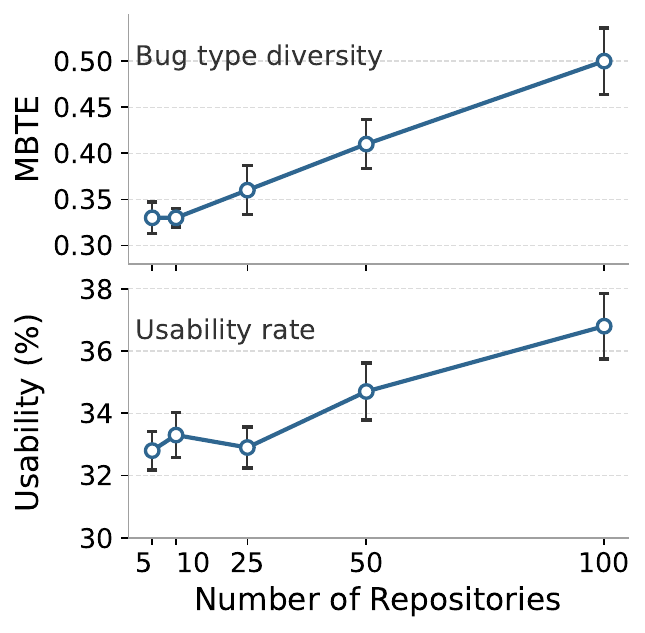
        }
        \captionof{figure}{Effect of repository coverage on bug type
        diversity and usability.}
        \label{fig:repository-diversity}
    \end{minipage}
    \hfill
    \begin{minipage}[t]{0.55\textwidth}
        \centering
        \includegraphics[height=2.0in]{
            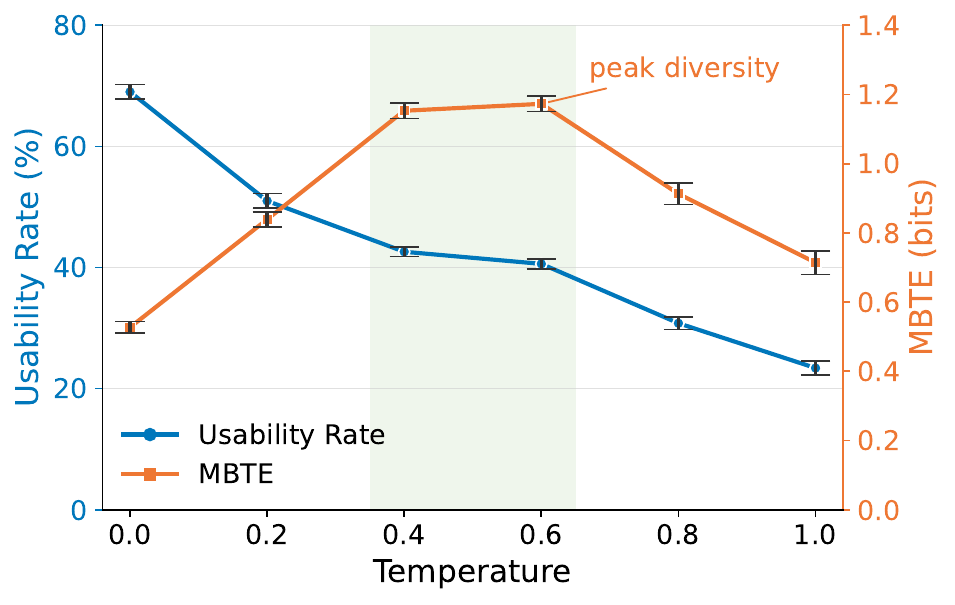
        }
        \captionof{figure}{Effect of sampling temperature on bug type diversity
        and usability.}
        \label{fig:temperature-diversity}
    \end{minipage}
    \vspace{-16pt}
\end{figure*}

\textbf{Broader repository coverage enables \approach to learn a more diverse bug distribution.}
To isolate the effect of repository coverage from training-set size, we train otherwise identical models using the same 2,000 training datapoints while varying only the number of repositories represented in the training data. For each model, we randomly select $N=50$ target files spanning all 12 repositories in SWE-bench Verified and generate ten independent candidates for each file with sampling temperature $t=0$, yielding 500 candidates per model. As shown in Figure~\ref{fig:repository-diversity}, MBTE increases as the training pool expands from 5 to 100 repositories, while usability remains relatively stable and improves modestly for the largest training pools. This result shows that both the diversity and usability of synthesized bugs increase as the number of training repositories grows.

\textbf{Moderate sampling temperatures maximize within-file bug diversity, while further increasing the temperature reduces both diversity and usability.}
We randomly select $N=50$ target files spanning all 12 repositories in SWE-bench Verified and generate ten independent candidates for each file at six temperatures,
$\{0.0, 0.2, 0.4, 0.6, 0.8, 1.0\}$, yielding 3,000 candidates in total. 

As shown in Figure~\ref{fig:temperature-diversity}, MBTE
increases from 0.527 at temperature 0.0 to 1.153 at 0.4 and peaks at 1.173 at 0.6, demonstrating that sampling enables the same target file to yield different bug types. However, usability decreases from 69.0\% at temperature
0.0 to 23.4\% at 1.0, while entropy also declines to 0.714 at the highest temperature. Moderate temperatures of 0.4--0.6 therefore provide the most favorable diversity--usability trade-off: they substantially expand the range of bugs generated from the same code context, whereas further increasing temperature primarily sacrifices usability without introducing additional diversity.

\begin{wrapfigure}{r}{0.43\textwidth}
   \centering
   \vspace{-10pt}
   \includegraphics[width=\linewidth]{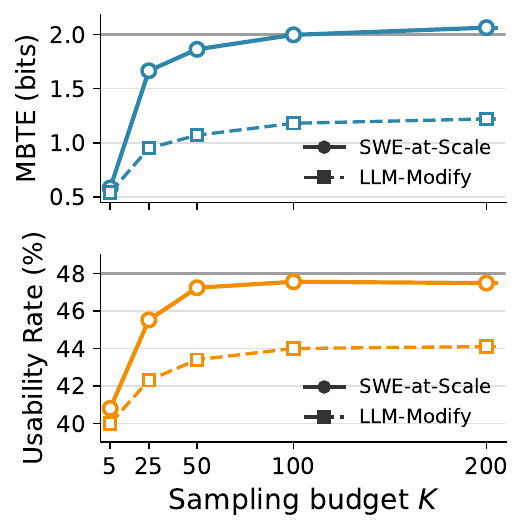}
   \caption{Sampling-budget scaling for \approach and LLM-Modify.}
   \label{fig:sampling-budget}
\end{wrapfigure}

\textbf{Larger sampling budgets continuously expand bug diversity while preserving synthesis usability.}
To study inference-time scaling, we fix both the target files and sampling temperature and progressively increase the sampling budget.
Specifically, we use the same 50 target files spanning 12 repositories, set the temperature to $t=0.6$, and evaluate nested budgets of $K\in\{5,25,50,100,200\}$ using prefixes of the same 205-sample pool. As shown in Figure~\ref{fig:sampling-budget}, \approach's MBTE increases from 0.583 at $K=5$ to 2.063 bits at $K=200$, while usability stabilizes at approximately 47\% beyond $K=50$. MBTE grows rapidly at smaller sampling budgets and begins to converge around $K=100$. Nevertheless, it continues to increase steadily up to $K=200$, indicating that additional sampling still expands bug-type diversity rather than reaching complete saturation. At the bug-patch level, 4,192 of 4,749 usable candidates (88.27\%) remain unique after normalized-diff deduplication at $K=200$.

In comparison, LLM-Modify with DS-V4-Flash reaches only 1.22 bits of MBTE and 44.1\% usability at $K=200$, with 61.74\% of its usable candidates remaining unique at the normalized-patch level. The substantially higher MBTE and unique-patch rate of \approach suggest that explicit training on reversed real-fix transformations enables the model to explore a broader space of bug types and concrete realizations as the sampling budget grows, whereas a general LLM without dedicated bug-inducing training exhibits substantially stronger repetition.

\subsection{Bug Complexity and Synthesis Cost}

\textbf{\approach synthesizes bugs that require larger code changes.}
As shown in Appendix Figure~\ref{fig:patch-structure}, \approach and
\approach+ modify 18.793 and 18.536 lines on average, respectively,
compared with 14.334 lines in SWE-bench Verified. The larger bug-inducing
patches suggest that \approach captures more complex code changes rather than
being limited to small local perturbations. We exclude LLM-Rewrite from this
comparison because its rewritten regions contain substantial code that
reconstructs the original functionality and therefore do not provide a
comparable measure of bug-inducing changes.
Figure~\ref{fig:token-cost} shows that the direct \approach path consumes 17.45k
tokens per instance on average, more than LLM-Modify (4.46--10.23k) but less
than LLM-Rewrite (34.92--62.82k). FeatAdd is an extreme outlier at 2.29M
tokens per instance, over two orders of magnitude above \approach. Thus,
\approach synthesizes relatively large bug-inducing patches while maintaining
substantially lower inference cost than rewrite-based synthesis.

\section{Related Work}

\subsection{Real-World Software Engineering Benchmarks}

Repository-level software engineering evaluation~\citep{chen2026performanceoptimizationbenchmarksreliablymeasuring} has rapidly evolved around SWE-bench, which transforms real-world GitHub issues into executable task instances for evaluating coding agents~\citep{chang2026test,zeng2025pruning,zeng2026dockerless,zeng2026glimprouter,gao2026swe,hu2026line,10.1145/3597503.3639085,10.1109/ICSE55347.2025.00012,10.1109/ASE63991.2025.00020,wang2026sweprunerselfadaptivecontextpruning,zhang2026paratempo,DBLP:journals/corr/abs-2508-13666}. SWE-bench~\citep{jimenez2024swebenchlanguagemodelsresolve} established repository-level issue resolution as an executable benchmark by pairing 2,294 real GitHub issues with their codebases and regression tests. SWE-rebench~\citep{badertdinov2025swerebenchautomatedpipelinetask} develops an automated pipeline that continuously extracts interactive tasks from GitHub and supports fresh, repository-disjoint evaluation to reduce contamination.
SWE-Bench Pro~\citep{deng2025swebenchproaiagents} targets substantially longer-horizon, enterprise-level work through 1,865 human-verified tasks drawn from public, held-out, and commercial repositories. DeepSWE~\citep{huang2026deepswemeasuringfrontiercoding} introduces 113 original, long-horizon SWE tasks across 91 open-source repositories and five programming languages, with newly authored reference solutions that are not merged into upstream repositories to reduce contamination from public development histories. SWE-Bench ProMax~\citep{shi2026swebenchpromaxbenchmarkingagents} instead evaluates large-scale, behavior-preserving refactoring with 170 expert-curated instances derived from real commits across seven languages.

However, naturally occurring task collections remain bounded by the pace of software development, repository availability, and the cost of task authoring and validation. \approach complements these efforts by learning from real software fixes to synthesize new executable task instances beyond the observed set of real-world issues.

\subsection{Synthetic Software Engineering Tasks}

To overcome the limited scale and coverage of naturally occurring SWE tasks~\citep{lyu2026practitionersbuildseagents, chen2026rethinkingvalueagentgeneratedtests, sun2026executinggeneratehidingexecution}, recent work~\citep{jain2025r2egymproceduralenvironmentshybrid, pan2025trainingsoftwareengineeringagents, lee2025gistifycodebaselevelunderstandingruntime} has explored scalable ways to construct executable software engineering tasks. 
SWE-smith~\citep{yang2025swesmith} automatically constructs executable environments for Python repositories and applies multiple bug-generation procedures, including LLM-based rewriting of code entities from their specifications rather than modifying their original implementations, thereby introducing implementation-level discrepancies that can naturally arise during software development. In contrast, SWE-Synth~\citep{pham2025swesynthsynthesizingverifiablebugfix} performs targeted LLM-based rewriting of existing code, followed by execution-based filtering, to generate verifiable bug-fix pairs and their intermediate repair trajectories. BugPilot~\citep{sonwane2025bugpilotcomplexbuggeneration} asks an agent to implement a feature and retains unintended test regressions as synthetic bugs, making bug creation resemble feature development rather than explicit mutation. 
SWE-Mirror~\citep{wang2025swemirrorscalingissueresolvingdatasets} abstracts the behavioral semantics of an existing issue and re-instantiates the same problem pattern in another functionally similar repository with a configured execution environment. 

Existing methods largely rely on inference-time prompting or source-issue transfer, so their synthesis diversity is constrained by the bug-generation knowledge already encoded in general LLMs or by the specific issues used as templates. In contrast, \approach converts real software fixes into explicit training supervision and learns a reusable bug-inducing distribution. This allows the learned bug patterns to be repeatedly instantiated in diverse forms across new code contexts, providing a more scalable way to expand the diversity of synthesized SWE tasks.

\section{Discussion}

\paragraph{Limitations and future directions.}
First, our current study focuses exclusively on Python repositories, and it remains unclear whether the learned reverse-transformation paradigm generalizes to other programming languages with different syntax, semantics, and software development practices. 
Consequently, realistic bugs that are not covered by the available test suite may be discarded, limiting the space of tasks that can be constructed from repositories with weak test coverage. 
Future work could investigate whether bug-inducing models can learn to operate across multiple programming languages, enabling scalable cross-language SWE task synthesis.

\paragraph{Conclusion.}
This paper presented \approach, a framework for scalable SWE task synthesis that learns bug injection by reversing real software fixes, rather than relying solely on the bug-generation capabilities of general LLMs. Our experiments show that \approach
produces realistic and diverse tasks that remain challenging for current
issue-resolution models. We further show that its learned bug distribution can
be explored at inference time: different bug types can be generated from the
same code context, and bug-type diversity continues to increase as the sampling
budget grows without fully saturating within 200 generations per file. 

\subsection*{AI Use Statement}
AI tools were used to assist with grammar checking of the manuscript and with writing code for data processing and visualization in the experiments.
\subsection*{Reproducibility Statement}
Our code and models are available at \url{https://github.com/cslsolow/SWE-at-scale} and \url{https://huggingface.co/SWE-at-Scale}.

\bibliography{iclr2027_conference}
\bibliographystyle{iclr2027_conference}

\newpage
\appendix
\setcounter{section}{1}

\section*{Appendix}

\subsection{LoRA Fine-Tuning Configuration}
We report the complete LoRA fine-tuning configuration used to train both \approach and \approach+. Table~\ref{tab:lora-config} summarizes the backbone, optimization hyperparameters, and training settings used in all experiments.

\begin{table}[H]
\centering
\small
\setlength{\tabcolsep}{5pt}
\begin{tabular}{ll}
\toprule
Hyperparameter & Value \\
\midrule
Base model & Qwen3.5-9B \\
Fine-tuning method & LoRA supervised fine-tuning \\
LoRA rank $r$ & 16 \\
LoRA scaling $\alpha$ & 32 \\
LoRA dropout & 0.05 \\
Target modules & All linear layers \\
Training examples & 13,407 \\
Maximum sequence length & 64,000 tokens \\
Number of epochs & 1 \\
Per-device batch size & 1 \\
Gradient accumulation steps & 8 \\
Effective batch size & 8 \\
Learning rate & $1\times10^{-4}$ \\
Learning-rate scheduler & Cosine \\
Warmup ratio & 0.03 \\
Weight decay & 0 \\
Maximum gradient norm & 1.0 \\
Numerical precision & bfloat16 \\
Gradient checkpointing & Enabled \\
\bottomrule
\end{tabular}
\caption{LoRA fine-tuning configuration used for both \approach and
\approach+.}
\label{tab:lora-config}
\end{table}

\subsection{Bug Complexity and Synthesis Cost}
\begin{figure*}[!htbp]
   \centering
   \begin{subfigure}[t]{0.49\textwidth}
      \centering
      \includegraphics[width=\linewidth]{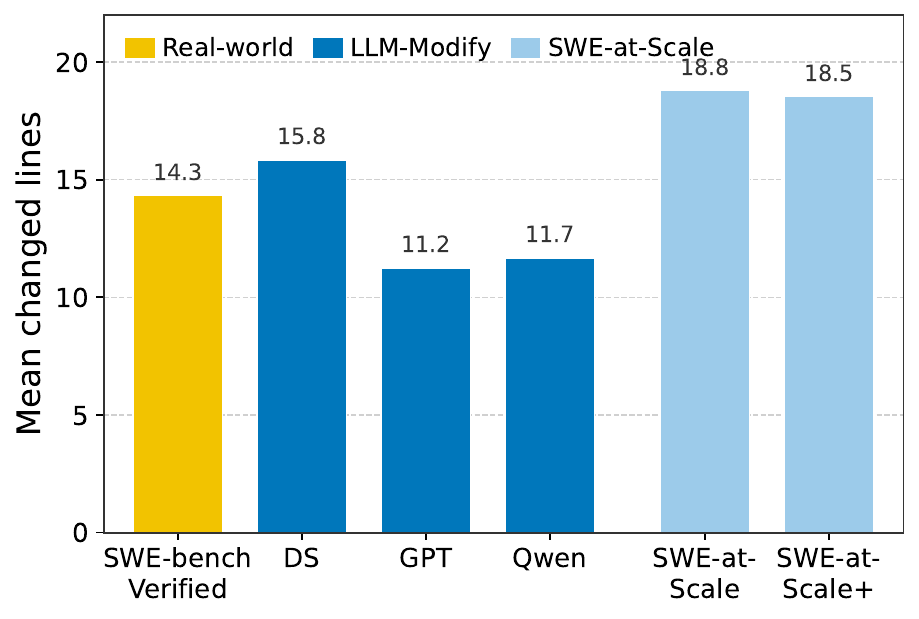}
      \caption{Mean changed lines per instance.}
      \label{fig:patch-structure}
   \end{subfigure}\hfill
   \begin{subfigure}[t]{0.49\textwidth}
      \centering
      \includegraphics[width=\linewidth]{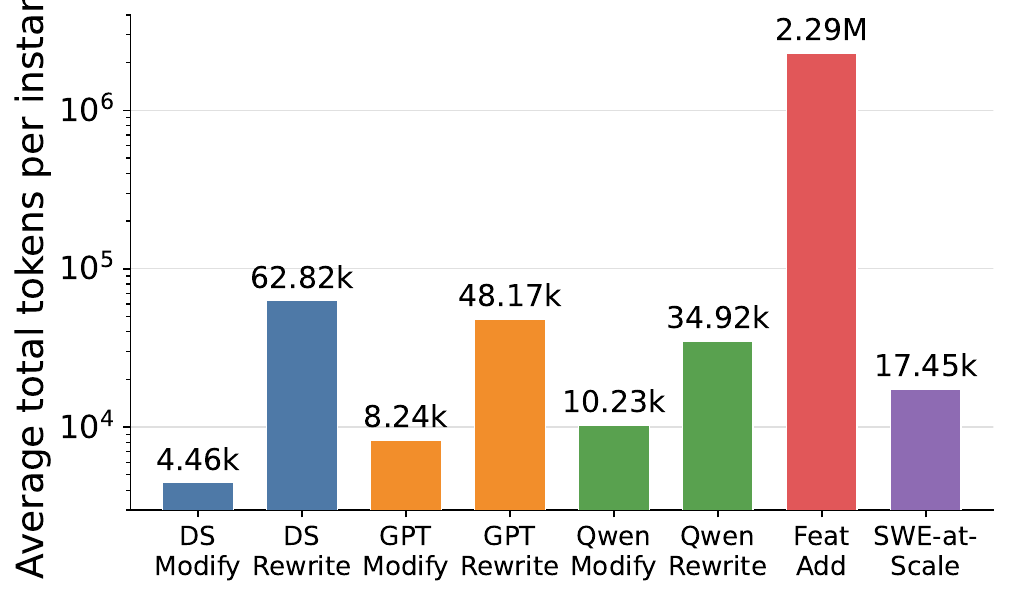}
      \caption{Average total tokens per instance.}
      \label{fig:token-cost}
   \end{subfigure}
   \caption{Mean changed lines and average total token consumption across the
   evaluated synthesis configurations.}
   \label{fig:patch-cost}
\end{figure*}

Figure~\ref{fig:patch-cost} reports the mean number of changed lines per synthesized instance and the average total token consumption required to produce one instance.

\end{document}